\documentclass[12pt]{article}
\usepackage{amsmath,amssymb,amscd,cite, hyperref}
\usepackage{color,soul,xcolor}
\definecolor{lightblue}{rgb}{.90,.95,1}
\newtheorem{thm}{Theorem}[section]
\newtheorem{prop}[thm]{Proposition}
\newtheorem{lem}[thm]{Lemma}
\newtheorem{cor}[thm]{Corollary}

\newtheorem{defi}[thm]{Definition}
\newcommand{\pf}{{\bf Proof. \ }}
\newcommand{\qed}{\hfill $\blacksquare$ \\}

\font\msbm=msbm10 at 12pt
\newcommand{\Z}{\mbox{\msbm Z}}

\date{}

\begin{document}

\title{ $\mathbb{Z}_{q}(\mathbb{Z}_{q}+u\mathbb{Z}_{q})$-Linear Skew Constacyclic Codes}

\author{Ahlem Melakhessou \thanks{ a.melakhessou@univ-batna2.dz, Department of Mathematics, Mostefa Ben Boula\"{i}d University (Batna2),
Batna, Algeria.}, Nuh Aydin \thanks{aydinn@kenyon.edu, Department of Mathematics and Statistics, Kenyon College, United States.} and Kenza Guenda \thanks{ University of Science and technology Houari Boumedien Algiers, Algeria, ken.guenda@gmail.com.}}
\maketitle

\begin{abstract}
 In this paper, we study skew constacyclic codes over the ring $\mathbb{Z}_{q}R$ where $R=\mathbb{Z}_{q}+u\mathbb{Z}_{q}$,
$q=p^{s}$ for a prime $p$ and  $u^{2}=0.$ We give the definition of these codes as subsets of the ring $\mathbb{Z}_{q}^{\alpha}R^{\beta}$. Some structural properties of the skew polynomial ring $ R[x,\theta]$ are discussed, where $ \theta$ is an automorphism of $R.$ We describe the generator polynomials of skew constacyclic codes over $ R $ and $\mathbb{Z}_{q}R.$ Using Gray images of skew constacyclic codes over $\mathbb{Z}_{q}R$ we obtained some new linear codes over $\mathbb{Z}_4$. Further, we have generalized these codes to double skew constacyclic codes over $\mathbb{Z}_{q}R$.
\end{abstract}
\section{Introduction}
Codes over finite rings have been known for several decades, but interest in these codes
increased substantially after the discovery that good non-linear binary codes can be constructed from codes over rings. Several methods have been introduced to produce certain types of linear codes with good algebraic structures and parameters.
Cyclic codes and their various generalizations such as constacyclic codes and quasi-cyclic (QC) codes have played a key role in this quest. One particularly useful generalization of cyclic codes has been the class of quasi-twisted (QT) codes that produced hundreds of new codes with best known parameters \cite{qtmain, qc1, qc2, qc3, qc4, qc5, qc6} recorded in the database \cite{database}. Yet another generalization of cyclic codes, called skew cyclic codes, were introduced in \cite{skew1} and they have been the subject of an increasing research activity over the past decade. This is due to their algebraic structure and  their  applications to DNA codes and quantum codes \cite{nabil,abhay, ezerman}. Skew constacyclic codes over various rings have been studied in \cite{abua, taher , ayats, rama, gao, fu, gursoy, qian, siap, yildiz} as a generalization of skew cyclic codes over finite fields.
 Recently, P. Li et al. \cite{li} gave the structure of $ (1+u)$-constacyclic codes over the ring $ \Z_{2}\Z_{2}[ u] $ and Aydogdu et al. \cite{ismail} studied
$\Z_{2}\Z_{2}[u]$-cyclic and constacyclic codes. Further, Jitman et al. \cite{jit} considered the
structure of skew constacyclic codes over finite chain rings. More recently A. Sharma and M. Bhaintwal studied skew cyclic codes over ring $\Z_{4}+u \Z_{4}$, where $ u^{2}=0.$

The aim of this paper is to introduce and study skew constacyclic codes over the ring  $\mathbb{Z}_{q}(\mathbb{Z}_{q}+u\mathbb{Z}_{q})$, where $q$ is a prime power and $u^2=0$.
 Some structural properties of the skew polynomial ring $ R[x,\theta]$ are discussed, where $ \theta$ is an automorphism of $R.$ We describe the generator polynomials of skew constacyclic codes over $ R $ and $\mathbb{Z}_{q}R.$ Using Gray images of skew constacyclic codes over $\mathbb{Z}_{q}R$ we obtained some new linear codes over $\mathbb{Z}_4$. Further, we  generalize these codes to double skew constacyclic codes over $\mathbb{Z}_{q}R$.

The paper is organized as follows. We first give some  basic results about the ring $R= \Z_{q}+u\Z_{q}$, where $ q=p^{s}$, $p$ is a prime and  $u^{2}=0$, and  linear codes over $\mathbb{Z}_{q}R$.
In Section 3, we construct the non-commutative ring $ R[ x;\theta]$, where  the structure of this ring depends on the elements of the commutative ring $R$ and an automorphism $ \theta $ of $R$.
In Section 4, we give some results on skew constacyclic codes over the ring $R$ and we determine  Gray images of skew constacyclic codes over $R$. In Section 5, we study the algebraic structure of skew constacyclic codes over the ring $\mathbb{Z}_{q}R.$ A necessary and sufficient condition for a skew constacyclic code over $\mathbb{Z}_{q}R$ to contain its dual is given, and we determine  Gray images of skew constacyclic codes over $\mathbb{Z}_{q}R$. These codes are then further generalized to double skew constacyclic codes. Finally, using Gray images of skew constacyclic codes over $\mathbb{Z}_{q}R$ we obtained some new linear codes over $\mathbb{Z}_4$.

\section{Preliminaries}

Let $( \alpha, \beta)$ denote $n=\alpha+2\beta $ where $ \alpha $ and $ \beta $ are positive integers. Consider the ring $ R= \mathbb{Z}_{q}+u\mathbb{Z}_{q}$, where $ q=p^{s}$, $p$ is a prime and $u^{2}=0$. The ring $R$ is isomorphic to the quotient ring $ \mathbb{Z}_{q}[ u ] / \left\langle u^{2} \right\rangle$. The ring $R$ is not a chain ring, whereas it is a local ring with the maximal ideal $\langle u, p \rangle$. Each element $r$ of $R$ can be expressed uniquely as
\begin{equation*}
r =a+ub, \text{ where } a, b\in \Z_{q}.
\end{equation*}
An element $ a+ub $ of $R$ is a unit if and only if $ a $ is unit. For a linear code $ C_{\beta} $ of length $ \beta $ over $ R $, its torsion $Tor( C_{\beta})$ and residue $Res( C_{\beta})$ codes are codes over $\Z_{q},$ defined as follows
\[
Tor( C_{\beta})= \{b \in \Z^{\beta}_{q}: \, ub \in C_{\beta}  \}
\]
and
\[
Res( C_{\beta})= \{a \in \Z^{\beta}_{q}:\, a+ub \in C_{\beta}  \text{ for some } b \in \Z^{\beta}_{q}\}.
\]
 \noindent Next we construct the ring
\[
\mathbb{Z}_{q}R = \lbrace (e,r): \, e \in \mathbb{Z}_{q}, r \in R \rbrace.
\]
The ring $\mathbb{Z}_{q}R $ is not an $R$-module under the operation of standard multiplication.
To make $\mathbb{Z}_{q}R $ an $R$-module, we follow the approach in \cite{taher} and  define the map
\[
\begin{array}{c}
\eta : R \rightarrow \mathbb{Z}_{q}\\
a+ub \mapsto a.
\end{array}
\]
Then, for any $d\in R$, we define the multiplication $ \star $ by
\[
d \star(e,r)=(\eta(d)e,dr).
\]
\noindent This multiplication can be naturally generalized to the ring $ \mathbb{Z}_{q}^{\alpha}R^{\beta}$ as follows.\\
For any $ d \in R $ and $ v=(e_{0}, e_{1}, \ldots , e_{\alpha-1}, r_{0}, r_{1}, \ldots , r_{\beta-1})\in \mathbb{Z}_{q}^{\alpha}R^{\beta}$ define
\[
dv=(\eta(d)e_{0}, \eta(d)e_{1}, \ldots , \eta(d)e_{\alpha-1}, dr_{0}, dr_{1}, \ldots , dr_{\beta-1}),
\]
where $( e_{0}, e_{1}, \ldots , e_{\alpha-1})\in \mathbb{Z}_{q}^{\alpha} $ and $( r_{0}, r_{1}, \ldots , r_{\beta-1}) \in R^{\beta} $.
\noindent The following results are analogous to the ones obtained in \cite{taher, ayats} for the ring $ \Z_{2}(\Z_{2}+u\Z_{2})$.
\begin{lem}
\label{Lem:jus}
The ring $ \Z^{\alpha}_{q}R^{\beta}$ is an $R$-module under the above definition.
\end{lem}
Lemma \ref{Lem:jus} allows us to give the next definition.
\begin{defi}
A non-empty subset $C$ of $ \Z^{\alpha}_{q}R^{\beta}$ is called a $\Z_{q}R$-linear code if it is an $R$-submodule of
$\Z^{\alpha}_{q}R^{\beta}$.
\end{defi}
We note that the ring $R^{\beta}$ is isomorphic to $ \Z^{2\beta}_{q} $ as an additive group. Hence, for some positive integers $k_{0}$, $k_{1}$ and $k_{2}$, any $ \Z_{q}R$-linear code $ C $ is isomorphic  to a group of the form
\[
\Z^{k_{0}}_{q} \times  \Z^{2k_{1}}_{q} \times  \Z^{k_{2}}_{q}.
\]
\begin{defi}
If $C\subseteq \Z^{\alpha}_{q} R^{\beta}$ is a $\Z_{q} R$-linear code, group isomorphic to $\Z^{k_{0}}_{q} \times  \Z^{2k_{1}}_{q} \times  \Z^{k_{2}}_{q}$, then $C$ is called a $\Z_{q} R$-additive code of type $(\alpha, \beta; k_{0}, k_{1}, k_{2})$, where $ k_{0} $, $ k_{1} $, and $ k_{2} $ are as defined above.
\end{defi}
\noindent The following results and definitions are  analogous to the ones obtained in \cite{taher}.\\

Let $ C $ be a $ \Z_{q}R$-linear code and  let $C_{\alpha}$ (respectively $ C_{\beta} $) be the canonical projection of $C$ on the first $\alpha$ (respectively on the last $\beta$) coordinates. Since the canonical projection is a linear map,  $C_{\alpha} $ and $C_{\beta}$
are linear codes over $\Z_{q}$ and over $R$ of length $\alpha$ and $\beta$, respectively. A code C is called separable if C is the direct product of $C_{\alpha}$  and $C_{\beta}$, i.e.,
 \[
 C=C_{\alpha}\times C_{\beta}.
\]
We introduce an inner product on $\Z_{q}^{\alpha}R^{\beta}$. For any two vectors
\[
  v =(v_{0}, \ldots , v_{\alpha-1}, v'_{0}, \ldots , v'_{\beta-1}), w =(w_{0}, \ldots , w_{\alpha-1}, w'_{0}, \ldots , w'_{\beta-1})\in \Z_{q}^{\alpha}R^{\beta}
 \]
let
\[
\langle v, w \rangle = u\sum \limits_{i=0}^{\alpha-1} v_{i}w_{i} + \sum \limits_{j=0}^{\beta-1} \acute{v}_{j}\acute{w}_{j}.
\]
Let $C$ be a $ \Z_{q}R$-linear code. The dual of $C$ is defined by

\[
C^{\perp}=\lbrace v\in \Z_{q}^{\alpha}R^{\beta} ;\,  \left\langle v, w \right\rangle=0, \forall w\in C \rbrace.
\]
If $C=C_{\alpha}\times C_{\beta}$ is separable, then

\begin{equation} \label{equation}
C^{\perp}=C^{\perp}_{\alpha}\times C^{\perp}_{\beta}.
\end{equation}

\section{Skew Polynomial Ring $ R[ x;\theta]  $}
In this section we construct the non-commutative ring $ R[ x;\theta]$. The structure of this ring depends on the elements of the commutative ring $R$ and an automorphism $\theta $ of $R$. Note that an automorphism $\theta$ of $R$ must fix every element of $\mathbb{Z}_q$, hence it satisfies $\theta(a+ub)=a+\theta(u)b$. Therefore, it is determined by its action on $u$. Let $\theta(u)=k+ud$, where $k$ is a non-unit in $\mathbb{Z}_q$, $k^{2} \equiv 0 \mod q$ and $2kd \equiv 0 \mod q$. Then, $\theta(a+ub)=a+\theta(u)b=(a+kb)+udb$ for all $ a + ub \in R $. Let $\theta$ be an automorphism of $R$ and let $m$ be its order.
The skew polynomial ring $ R[x;\theta] $ is the set of polynomials over $R$ where the addition of these polynomials is defined in the usual way while multiplication $\ast$ is defined using the distributive law and the rule
\[
x\ast a=\theta(a)x.
\]
The set $R[ x;\theta] $ with respect to addition and multiplication defined above form a non-commutative ring called the skew polynomial ring.
 An element $ g(x)\in  R[ x;\theta]$ is said to be a right divisor (resp. left divisor) of $f(x)$ if there exists $ q(x) \in  R[ x;\theta] $ such that
\[
f(x)= q(x) \ast g(x)  \quad (\text{ resp. } f(x)= g(x)\ast q(x)).
\]
In this case, $f(x)$ is called a left multiple (resp. right multiple) of $g(x)$.
\begin{lem} \cite[Lemma 1]{amit}
Let $ f(x) $, $ g(x) \in R[ x;\theta]$ be such that the leading coefficient of $ g(x) $ is a unit. Then there exist $q(x)  $, $ r(x)\in R[ x;\theta] $ such that
\[
f(x) = q(x) \ast g(x)+r(x),  \text{ where }  r(x)=0 \text{ or } \deg(r(x)) < \deg(g(x)).
\]
\end{lem}

\begin{defi}
A polynomial $ f(x) \in R[ x;\theta] $ is said to be a central polynomial if
\[
 f(x)\ast r(x) =  r(x) \ast f(x) \qquad \text{ for all  } r(x) \in R[ x;\theta]
\]
\end{defi}

\begin{thm}
The center $Z(R[x;\theta])$ of $ R[ x;\theta] $ is the set $R[x^{m}]$, where $m$ is the order of $ \theta \in Aut(R)$.
\end{thm}
\pf
An automorphism $\theta$ of $R$ must fix every element of $\Z_{q}$. Since $m$ is
the order of the automorphism $\theta$, for any $ a \in R$, we have
\[
x^{mi} \ast a=(\theta^{m})^{i}(a)x^{mi}=a \ast x^{mi}.
\]
Thus, $x^{mi}$ is in the center $ Z(R[x;\theta])$ of $ R[ x;\theta] $. This implies that
\[
f(x) =\xi_{0} +\xi_{1}x^{m}+ \xi_{2}x^{2m}+ \ldots +\xi_{l}x^{lm},
\]
where $ \xi_{i} \in R$, is in the center. Conversely, let $  f(x) =\xi_{0} +\xi_{1}x+ \xi_{2}x^{2}+ \ldots +\xi_{k}x^{k} $, where $ \xi_{i} \in R$ for $ i=0, \ldots, k $, be in the center of $ R[ x; \theta] $. Then $a \ast f(x) = f(x)\ast a $ for any $ a \in R $. Hence,
\[
\begin{array}{cccl}
   f(x) \ast a&= & a \ast f(x)\\
(\xi_{0} +\xi_{1}x+ \xi_{2}x^{2}+ \ldots +\xi_{k}x^{k}) \ast a &=& a \ast  (\xi_{0} +\xi_{1}x+ \xi_{2}x^{2}+ \ldots +\xi_{k}x^{k}) \\
a \xi_{0} + \xi_{1} \theta(a)x+ \xi_{2} \theta^{2}(a)x^{2}+\ldots+\xi_{k} \theta^{k}(a)x^{k} &=& a \xi_{0} +  a \xi_{1} x + a \xi_{2} x^{2} + \ldots+ a \xi_{k} x^{k}.
\end{array}
\]

\noindent Therefore, for each $i$ we have $  \xi_{i} \theta^{i}(a) = \xi_{i}a$. Since this is true for all $a\in R$, $\theta^{i}(a)=a$ which implies that $m|i$. This means 
\[
f(x) = \xi_{0} +\xi_{1}x^{m} + \xi_{2}x^{2m}+ \ldots + \xi_{l}x^{lm}.
\]
Thus, any element of center is in $R[x^{m}].$ This completes the proof. \qed

\noindent The following result easily follows for a non zero element $\lambda_{0}+u\lambda_{1} \in R$.
\begin{cor}
\label{side}
Let $ f(x)=x^{\beta}-1 $. Then $f(x) \in Z(R[ x;\theta])$ if and only if $ m\mid\beta.$ Further, $ x^{\beta}-(\lambda_{0}+u\lambda_{1})\in Z(R[ x;\theta]) $
if and only if $ m \mid \beta$ and $ (\lambda_{0}+u\lambda_{1})$ is fixed by $ \theta.$
\end{cor}

\section{Skew $ (\lambda_{0}+u\lambda_{1})- $Constacyclic Codes over $ R $}
To study skew constacyclic codes over $ R $, we first consider some structural properties of $ R[ x;\theta]/ \langle x^{\beta}-(\lambda_{0}+u\lambda_{1}) \rangle $. The Corollary \ref{side}, shows that the polynomial $ (x^{\beta}-(\lambda_{0}+u\lambda_{1})) $ is in the center $Z(R[ x; \theta])$ of the ring $ R[x; \theta] $, hence generates a two-sided ideal if and only if $ m \mid \beta$ and $ (\lambda_{0}+u\lambda_{1})$ is fixed by $ \theta.$ Therefore, in this case $ R[ x; \theta]/ \langle x^{\beta}-(\lambda_{0}+u\lambda_{1}) \rangle  $ is a well-defined residue class ring. If $m \nmid \beta$, then the quotient space $R[ x; \theta]/ \langle x^{\beta}-(\lambda_{0}+u\lambda_{1}) \rangle $ which is not necessarily a ring is a left $ R[ x; \theta]$-module with multiplication defined by
\[
r(x) \ast (f(x)+( x^{\beta}-(\lambda_{0}+u\lambda_{1}))=r(x) \ast f(x)+(x^{\beta}-(\lambda_{0}+u\lambda_{1})),
\]
for any $ r(x) $, $ f(x)\in R[ x; \theta]$. Next we  define the skew $(\lambda_{0}+u\lambda_{1})$-constacyclic codes over the ring $ R $.
\begin{defi}
A subset $ C_{\beta} $ of $ R^{\beta} $ is called a skew $(\lambda_{0}+u\lambda_{1})$-constacyclic code of length $ \beta $ if
the following two conditions hold

(i) $ C_{\beta} $ is an $R$-submodule of $R^{\beta}.$

(ii) \[((\lambda_{0}+u \lambda_{1})\theta(c_{\beta-1}), \theta(c_{0}), \ldots ,\theta(c_{\beta-2})) \in C_{\beta},
\] whenever
\[
(c_{0}, c_{1}, \ldots , c_{\beta-1}) \in C_{\beta}.
\]

\end{defi}
In particular, if $ \lambda_{0}+u \lambda_{1}=1,$ then $C_{\beta}$ is a skew cyclic code over $R$, and we have the classical cyclic codes when $\theta=Id $ and $ \lambda_{0}+u \lambda_{1}=1$.
In polynomial representation, a codeword $ (c_{0}, c_{1}, \ldots , c_{\beta-1} )$ of a skew  $(\lambda_{0}+u\lambda_{1})$-constacyclic code is the vector of coefficients of the corresponding polynomial
\begin{equation*}
c_{0}+c_{1}x+ \ldots +c_{\beta-1}x^{\beta-1} \in R[ x; \theta]/ \langle x^{\beta}-(\lambda_{0}+u \lambda_{1}) \rangle.
\end{equation*}
\begin{lem}
\label{dualskewcon}
Let $C_{\beta} $ be a skew $( \lambda_{0}+u\lambda_{1}) $-constacyclic code of length $\beta$ over $ R $, where $ \beta$ is a multiple of the order $m$ of the automorphism $\theta$. Then the dual code $C_{\beta}^\bot $ for $C_{\beta}$ is a skew $ ( \lambda_{0}+u\lambda_{1})^{-1}$-constacyclic code of length $\beta$ over $ R $.
\end{lem}
\pf
Let $ v'=(v'_{0}, v'_{1}, \ldots , v'_{\beta-1}) \in C_{\beta}$ and $ w'=(w'_{0}, w'_{1}, \ldots , w'_{\beta-1})\in C_{\beta}^{\bot}$. Since
\[
\left( ( \lambda_{0}+u\lambda_{1})\theta^{\beta-1}(v'_{1}), ( \lambda_{0}+u\lambda_{1})\theta^{\beta-1}(v'_{2}), \ldots , ( \lambda_{0}+u\lambda_{1})\theta^{\beta-1}( v'_{\beta-1}),\theta^{\beta-1}(v'_{0})\right) \in C_{\beta},
\]
we have
\[
\begin{array}{ccl}
0&=&\langle ( ( \lambda_{0}+u\lambda_{1})\theta^{\beta-1}(  v'_{1}), ( \lambda_{0}+u\lambda_{1})\theta^{\beta-1}( v'_{2}), \ldots , ( \lambda_{0}+u\lambda_{1})\theta^{\beta-1}(v'_{\beta-1}),\theta^{\beta-1}(v'_{0})), w' \rangle\\
&=&\langle (( \lambda_{0}+u\lambda_{1}) \theta^{\beta-1}(v'_{1}), \ldots , ( \lambda_{0}+u\lambda_{1})\theta^{\beta-1}(v'_{\beta-1}), \theta^{\beta-1}( v'_{0})), (w'_{0}, \ldots , w'_{\beta-1}) \rangle\\
&=&( \lambda_{0}+u\lambda_{1})\langle ( \theta^{\beta-1}( v'_{1}), \ldots ,\theta^{\beta-1}(  v'_{\beta-1}), \theta^{\beta-1}(( \lambda_{0}+u\lambda_{1})^{-1}  v'_{0})), (w'_{0}, w'_{1}, \ldots , w'_{\beta-1}) \rangle\\
&=&( \lambda_{0}+u\lambda_{1})\left( \theta^{\beta-1}(( \lambda_{0}+u\lambda_{1})^{-1} v'_{0})w'_{\beta-1}+\sum \limits_{j=1}^{\beta-1}\theta^{\beta-1}(v'_{j})w'_{j-1}\right).
\end{array}
\]
As $\beta$ is a multiple of the order of $ \theta $ and $ (\lambda_{0}+u \lambda_{1})^{-1} $ is fixed by $ \theta $, it follows that
\[
\begin{array}{ccl}
0&=&\theta( ( \lambda_{0}+u\lambda_{1})( \theta^{\beta-1}(( \lambda_{0}+u\lambda_{1})^{-1} v'_{0})w'_{\beta-1}+\sum \limits_{j=1}^{\beta-1}\theta^{\beta-1}(v'_{j})w'_{j-1})) \\
&=&( \lambda_{0}+u\lambda_{1})\left(  v'_{0}\theta(( \lambda_{0}+u\lambda_{1})^{-1} w'_{\beta-1})+\sum \limits_{j=1}^{\beta-1}v'_{j}\theta(w'_{j-1})\right) \\
&=&( \lambda_{0}+u\lambda_{1}) \langle ( \theta( ( \lambda_{0}+u\lambda_{1})^{-1} w'_{\beta-1}),\theta(  w'_{0}), \ldots , \theta( w'_{\beta-2})), (v'_{0}, v'_{1}, \ldots , v'_{\beta-1}) \rangle.
\end{array}
\]
Therefore,
\[
(\theta((\lambda_{0}+u \lambda_{1})^{-1}w'_{\beta-1}), \theta(w'_{0}), \ldots , \theta(w'_{\beta-2})) \in C_{\beta}^{\bot}.
\]
\qed\\
We are now ready to consider the  generator polynomials of  skew constacyclic codes over $R$.

\begin{thm}
\label{lef}
A code $C_{\beta} $ of length $ \beta $ over $R$ is a skew $ ( \lambda_{0}+u\lambda_{1})$-constacyclic code if and only if $ C_{\beta}$ is a left $ R[x; \theta]$-submodule of $R[x; \theta]/\left\langle x^{\beta}-( \lambda_{0}+u\lambda_{1})\right\rangle $.
\end{thm}
\pf
Let $ c(x)=c_{0}+ c_{1} x+ \ldots+ c_{\beta-1}x^{\beta-1} \in C_{\beta}$, then
\[
\begin{array}{ccl}
x \ast c(x)&=&  (\lambda_{0}+u\lambda_{1})\theta(c_{\beta-1})+\theta(c_{0})x+ \ldots +\theta(c_{\beta-2})x^{\beta-1}\\
&=& ((\lambda_{0}+u\lambda_{1})\theta(c_{\beta-1}), \theta(c_{0}), \ldots , \theta(c_{\beta-2})) \in C_{\beta}.
\end{array}
\]
By iteration and linearity one obtains $ r(x) \ast c(x) \in C_{\beta}$, for all $ r(x) \in  R[x,\theta]/\langle x^{\beta}-(\lambda_{0}+u\lambda_{1})\rangle.$ This shows that $  C_{\beta} $ is a left $ R[x; \theta]$-submodule of $ R[x; \theta]/\left\langle x^{\beta}-( \lambda_{0}+u\lambda_{1})\right\rangle $.
Conversely, suppose that $ C_{\beta}$ is a left $R[x; \theta]$-submodule of $R[x; \theta]/\left\langle x^{\beta}-( \lambda_{0}+u\lambda_{1})\right\rangle $, then we have that $x \ast c(x) \in  C_{\beta}$. Thus, $ C_{\beta}$ is skew $ (\lambda_{0}+u\lambda_{1})$-constacyclic code.
\qed

\noindent The proofs of the next two theorems are analogous to  the proofs of Theorem 4 and Theorem 5 of \cite{amit} given for the ring $\Z_{4}+u\Z_{4}$, therefore we omit them.
\begin{thm}\cite[Theorem 4]{amit}
If $ C_{\beta} $ is a skew $ ( \lambda_{0}+u\lambda_{1})$-constacyclic code of length $ \beta $ over $ R $ containing a minimum degree polynomial $ g_{\beta}(x) $ whose leading coefficient is a unit, then $ C_{\beta} $ is a free code such that $ C_{\beta}=\left\langle g_{\beta}(x)\right\rangle  $ and $ g_{\beta}(x)\vert x^{\beta}-( \lambda_{0}+u\lambda_{1}).$ Further, $ C_{\beta}$ has a basis
\[
\{g_{\beta}(x), x \ast g_{\beta}(x), \ldots , x^{\beta-deg(g_{\beta}(x))-1} \ast g_{\beta}(x)\}
\]
and $ \vert C_{\beta}\vert= \vert R \vert^{\beta- deg(g_{\beta}(x))}$.
\end{thm}
The converse of the above theorem is given by the following.
\begin{thm}\cite[Theorem 5]{amit}
Let $ C_{\beta} $ be a free, principally generated skew $ ( \lambda_{0}+u\lambda_{1})$-constacyclic code of length $ \beta $ over $ R$. Then there exists a minimal degree polynomial $g_{\beta}(x) \in   C_{\beta} $ having its leading coefficient a unit such that $ C_{\beta}=\left\langle g_{\beta}(x)\right\rangle  $ and $ g_{\beta}(x) \vert x^{\beta}-( \lambda_{0}+u\lambda_{1}).$
\end{thm}

\subsection{ The Gray Images of  Skew Constacyclic Codes over $R$}

The purpose of this section is to investigate the Gray images of  skew constacyclic codes over $R$. We start with the following definition.
\begin{defi}
\label{twisted}
Let $ C$ be a linear code over $\Z_{q}$ of length $ n =Nl $ and let $ \lambda \in \Z^{*}_{q} $. If for any codeword
\[
(c_{0,0}, c_{0,1}, \ldots ,c_{0,l-1}, \ldots , c_{N-1,0}, c_{N-1,1}, \ldots ,c_{N-1,l-1}) \in C,
\]
we have that
\[
( \lambda c_{N-1,0}, \lambda c_{N-1,1}, \ldots , \lambda c_{N-1,l-1}, \ldots ,c_{0,0}, c_{0,1}, \ldots , c_{0,l-1}) \in C,
\]
then we say that $C$ is a $\lambda$-quasi-twisted (QT) code of length $n$. If $l$ is the least
positive integer satisfying $ n=Nl $, then $C$ is said to be a $\lambda$-quasi-twisted
code with index $l$.
\end{defi}
Define the Gray map $\Psi$ by

\[
\begin{array}{c}
\Psi :R ^{\beta} \rightarrow \Z^{2\beta}_{q}\\
(r_{0}, r_{1}, \ldots , r_{\beta-1}) \mapsto (b_{0}, b_{1}, \ldots , b_{\beta-1}, a_{0}+b_{0}, a_{1}+b_{1}, \ldots , a_{\beta-1}+b_{\beta-1})
\end{array}
\]
where $ r_{i}=a_{i}+ub_{i} \in R$, $ a_{i}, b_{i} \in \Z_{q} \text{ for } i=0, 1, \ldots, \beta-1$. We will use this map to obtain codes over $\Z_{q}$. The next proposition follows from the definition  of QT codes.

\begin{prop}
If $ C_{\beta} $ is a skew $ ( \lambda_{0}+u\lambda_{1}) $-constacyclic code of length $\beta$ over $R$, then $ \Psi(C_{\beta})$ is a QT  code of index $2$ and length $ 2\beta $ over $ \Z_q .$
\end{prop}
\pf
Let $ C_{\beta}$ be a skew constacyclic code in $ R[x; \theta]/ \langle x^{\beta}-(\lambda_{0}+u\lambda_{1})\rangle$ and let
\[
( b_{0}, b_{1}, \ldots , b_{\beta-1}, a_{0}+b_{0}, a_{1}+b_{1}, \ldots , a_{\beta-1}+b_{\beta-1})\in \Psi(C_{\beta}).
\]
Then there is a codeword $ (  a_{0}+ub_{0}, a_{1}+ub_{1}, \ldots , a_{\beta-1}+ub_{\beta-1})\in C_{\beta} $ such that

\[
\Psi(  a_{0}+ub_{0}, a_{1}+ub_{1}, \ldots , a_{\beta-1}+ub_{\beta-1})= (b_{0}, b_{1}, \ldots , b_{\beta-1}, a_{0}+b_{0}, a_{1}+b_{1}, \ldots , a_{\beta-1}+b_{\beta-1}).
\]
Since $ C_{\beta}$ is skew $(\lambda_{0}+u\lambda_{1})$-constacyclic, we have

\[
( (\lambda_{0}+u\lambda_{1}) \theta(r_{\beta-1}), \theta(r_{0}), \ldots , \theta(r_{\beta-2}))\in C_{\beta}
\]
and

\[
\begin{array}{ccl}
& &\Psi((\lambda_{0}+u\lambda_{1})\theta(a_{\beta-1}+ub_{\beta-1}), \theta(a_{0}+ub_{0}), \ldots ,  \theta(a_{\beta-2}+ub_{\beta-2}))\\
&=&\Psi((\lambda_{0}+u\lambda_{1})(a_{\beta-1}+kb_{\beta-1}+udb_{\beta-1}), a_{0}+kb_{0}+udb_{0}, \ldots , a_{\beta-2}+kb_{\beta-2}+udb_{\beta-2})\\
 &=&\Psi((\lambda_{0}(a_{\beta-1}+kb_{\beta-1})+u(\lambda_{1}a_{\beta-1}+(\lambda_{0}d+\lambda_{1}k)b_{\beta-1}), a_{0}+kb_{0}+udb_{0}, \ldots, a_{\beta-2}+kb_{\beta-2}+udb_{\beta-2}) \\
 &=&(\lambda_{1}a_{\beta-1}+(\lambda_{0}d+\lambda_{1}k)b_{\beta-1},db_{0} \ldots , db_{\beta-2},
(\lambda_{0}+\lambda_{1})a_{\beta-1}+((k+d)\lambda_{0}+k\lambda_{1}) b_{\beta-1},a_{0}+(k+d)b_{0} ,\ldots ,\\
&& a_{\beta-2}+(k+d)b_{\beta-2}).
\end{array}
\]
Hence, $ \Psi(C_{\beta}) $ is $ (\lambda_{0}+u\lambda_{1}) $-QT with index 2 of length $2\beta $. \qed

\section{$\mathbb{Z}_{q}R$ -Linear Skew $(\lambda_{0}+u\lambda_{1})- $Constacyclic Codes}
In this section, we study skew $(\lambda_{0}+u\lambda_{1})$-constacyclic codes over the ring $ \Z_{q}R.$

\begin{defi}
Let  $ \theta $ be an automorphism of $R$. A linear code $C$ over $\Z_{q}^{\alpha}R^{\beta}$ is called skew constacyclic if $C$ satisfies the following two conditions.

(i) $C$ is an $R$-submodule of $\Z_{q}^{\alpha}R^{\beta}$,

(ii)
\[
 \ (e_{\alpha-1} , e_{0}
, \ldots , e_{\alpha-2}, ( \lambda_{0}+u\lambda_{1} )\theta( r_{\beta-1}), \theta ( r_{0})
, \ldots , \theta ( r_{\beta-2}) ) \in C.
\]
whenever
\[
( e_{0}, e_{1}, \ldots , e_{\alpha-1}, r_{0}, r_{1}, \ldots , r_{\beta-1}) \in C
\]

\end{defi}

In polynomial representation, each codeword $c= (e_{0}, e_{1}, \ldots, e_{\alpha-1},r_{0},r_{1}, \ldots, r_{\beta-1}  ) $ of a skew constacyclic code can be represented by a pair of polynomials

 \begin{equation*}
c(x)=\left(
\begin{array}{c}
e_{0}+ e_{1}x+ \ldots + e_{\alpha-1}x^{\alpha-1},\\
r_{0}+ r_{1}x+ \ldots +r_{\beta-1}x^{\beta-1}
\end{array}
\right)
=(e(x), r(x))\in   \Z_{q}[x ]/ \langle x^{\alpha}-1 \rangle  \times R[x; \theta]/ \langle x^{\beta}-(\lambda_{0}+u\lambda_{1}) \rangle.
\end{equation*}
Let $ h(x)=h_{0}+ h_{1}x+ \ldots + h_{t}x^{t} \in R \ [ x; \theta ] $ and let $ (f(x), g(x) )\in  \Z_{q}[x ]/ \langle x^{\alpha}-1 \rangle  \times R[x; \theta]/ \langle x^{\beta}-(\lambda_{0}+u\lambda_{1})\rangle$. The multiplication is defined by the basic rule
\[
h(x) (f(x), g(x) )=(\eta(h(x))f(x), h(x) \ast g(x)),
\]
where $\eta(h(x))=\eta( h_{0})+\eta( h_{1})x+\ldots+\eta( h_{t})x^{t}.$

\begin{lem}
A code $ C $ of length $(\alpha, \beta )$ over $ \Z_{q}R $ is a skew $ (\lambda_{0}+u\lambda_{1})$-constacyclic code
 if and only if $ C$ is left $R[x; \theta ]$-submodule of $ \Z_{q}[x] / \langle x^{\alpha}-1 \rangle  \times R[x;\theta] / \langle x^{\beta}-(\lambda_{0}+u \lambda_{1}) \rangle $.
\end{lem}

\pf
Let $ c \in C $, where $ C $ is a skew $ (\lambda_{0}+u\lambda_{1})$-constacyclic code. We denote by $c(x) = (e(x), r(x))$ the associated polynomial of $c$. As $ x  c(x) $ is a skew constacyclic shift of $ c$, $ x c(x)\in C $. Then, by linearity of $ C$, $r(x) c(x) \in C $ for any $ r(x) \in R[x; \theta].$ Thus $ C $ is left $ R[x; \theta]$-submodule of $ \Z_{q}[x] / \langle x^{\alpha}-1 \rangle  \times R[x;\theta] / \langle x^{\beta}-(\lambda_{0}+u \lambda_{1}) \rangle $.
Conversely, suppose that $ C$ is a left $R[x; \theta]$-submodule of $ \Z_{q}[x] / \langle x^{\alpha}-1 \rangle \times R[x;\theta] / \langle x^{\beta}-(\lambda_{0}+u \lambda_{1}) \rangle $, then we have that $x c(x) \in C$. Thus, $ C$ is a skew $ (\lambda_{0}+u\lambda_{1})$-constacyclic code. \qed

 \begin{thm}
\label{sc}
Let $ C $ be a linear code over $ \Z_{q}R $ of length $ (\alpha,\beta) $, and let $C=C_{\alpha} \times C_{\beta}$, where $C_{\alpha} $ is linear code over $ \Z_q $ of length $ \alpha $ and $ C_{\beta} $ is linear code over $ R $ of length $ \beta $. Then $ C $ is a skew $ (\lambda_{0}+u\lambda_{1})$-constacyclic code with respect to the automorphism $ \theta $ if and only if $C_{\alpha} $ is a cyclic code over $ \Z_{q}$ and $C_{\beta} $ is a skew $ (\lambda_{0}+u\lambda_{1})$-constacyclic code over $ R$ with respect to the automorphism $\theta$.
\end{thm}

\pf
Let $ (e_{0}, e_{1}, \ldots , e_{\alpha-1}) \in C_{\alpha}$ and let $ (r_{0}, r_{1}, \ldots , r_{\beta-1}) \in C_{\beta}$. If $ C $ is a skew constacyclic code, then
\[
\begin{array}{ccc}
( e_{\alpha-1}, e_{0}, \ldots , e_{\alpha-2}, (\lambda_{0}+u\lambda_{1} )\theta(r_{\beta-1}),\theta(r_{0}), \ldots ,\theta(r_{\beta-2}))\in C,
\end{array}
\]
which implies that
\[
(e_{\alpha-1}, e_{0}, \ldots ,e_{\alpha-2}) \in C_{\alpha}
\]
and
\[
  (( \lambda_{0}+u\lambda_{1} )\theta(r_{\beta-1}), \theta(r_{0}), \ldots , \theta(r_{\beta-2})) \in C_{\beta}.
\]
 Hence, $C_{\alpha}$ is a cyclic code over $ \Z_{q} $ and $ C_{\beta}$ is a skew $(\lambda_{0}+u\lambda_{1})$-constacyclic code over $ R $ with respect to the automorphism  $ \theta $.

 On the other hand, suppose that $C_{\alpha}$ is a cyclic code over $ \Z_q $ and $C_{\beta} $ is a skew constacyclic code over $ R $.
Note that
\[
( e_{\alpha-1}, e_{0}, \ldots , e_{\alpha-2}) \in C_{\alpha}
\]
and
\[
 ((\lambda_{0}+u\lambda_{1})\theta(r_{\beta-1}), \theta(r_{0}), \ldots , \theta(r_{\beta-2}) )\in C_{\beta},
\]
 so $C$ is a skew constacyclic code over $ \Z_{q}R $.
\qed

\begin{cor}
\label{dualZ}
Let $C=C_{\alpha}\times C_{\beta}$ be a skew $ (\lambda_{0}+u \lambda_{1})$-constacyclic code over $ \Z_{q}R $, where $ \beta$ is a multiple of the order $m$ of the automorphism $\theta$. Then the dual code $ C^\bot=C^\bot_{\alpha}\times C^\bot_{\beta} $ of $ C $ is a skew $ (\lambda_{0}+u \lambda_{1})^{-1}$-constacyclic code over $ \Z_{q}R $.
\end{cor}

\pf
From Equation (\ref{equation}), we have $ C^{\bot}=C^{\bot}_{\alpha}\times C^{\bot}_{\beta} $.
Clearly, if $ C_{\alpha} $ is a cyclic code over $ \Z_{q} $ then $C^\bot_{\alpha}  $ is also a cyclic code over $ \Z_{q} $.
Moreover, from Lemma (\ref{dualskewcon}), we have $ C^{\perp}_{\beta} $ is a skew $ (\lambda_{0}+u \lambda_{1})^{-1}$-constacyclic code over $ R $. Hence the dual code $ C^{\bot} $ is skew $ (\lambda_{0}+u \lambda_{1})^{-1}$-constacyclic over $\Z_{q} R.$
\qed\\
If $\lambda_{0} +u \lambda_{1}=1,$ then $ C $ is called a skew cyclic code, and if $ \theta=Id $, then the code is just constacyclic.

\begin{cor}
\label{sc}
Let $ C $ be a linear code over $ \Z_{q}R $ of length $ (\alpha, \beta) $, and let $C=C_{\alpha} \times C_{\beta}$, where $C_{\alpha} $ is a linear code over $ \Z_q $ of length $ \alpha $ and $ C_{\beta} $ is a linear code over $ R $ of length $ \beta $. Then $ C $ is a skew cyclic code with respect to the automorphism $ \theta $ if and only if $C_{\alpha} $ is  a cyclic code over $ \Z_{q}$ and $C_{\beta}$ is a skew cyclic code over $ R$ with respect to the automorphism $ \theta $.
\end{cor}

\pf
Let $ (e_{0}, e_{1}, \ldots , e_{\alpha-1})\in C_{\alpha}$ and let $ (r_{0}, r_{1}, \ldots , r_{\beta-1}) \in C_{\beta}$. If $C$ is a skew cyclic code, then
\[
\begin{array}{ccc}
(e_{\alpha-1}, e_{0}, \ldots , e_{\alpha-2}, \theta(r_{\beta-1}), \theta(r_{0}), \ldots ,\theta(r_{\beta-2}))\in C,
\end{array}
\]
which implies that
\[
( e_{\alpha-1}, e_{0}, \ldots , e_{\alpha-2}) \in C_{\alpha},
\]
and
\[
 (\theta(r_{\beta-1}), \theta(r_{0}), \ldots , \theta(r_{\beta-2})) \in C_{\beta}.
\]
 Hence, $ C_{\alpha}  $ is a cyclic code over $ \Z_{q} $ and $ C_{\beta}$ is a skew cyclic code over $ R $ with respect to the automorphism $\theta $.
On the other hand, suppose that $C_{\alpha} $ is a cyclic code over $ \Z_{q} $ and $C_{\beta} $ is a skew cyclic code over $ R $.
Note that
\[
\begin{array}{cccc}
(e_{\alpha-1}, e_{0}, \ldots , e_{\alpha-2}) \in C_{\alpha}
\end{array}
\]
and
\[
(\theta(r_{\beta-1}), \theta(r_{0}), \ldots , \theta(r_{\beta-2}) )\in C_{\beta},
\]
 so $C$ is a skew cyclic code over $ \Z_{q}R $. \qed

\begin{thm}
Let $C=C_{\alpha} \times C_{\beta}$ be a linear code over $ \Z_{q}R $ of length $ (\alpha, \beta) $, and let $ C_{\alpha} $ be a principally generated free cyclic code of length $ \alpha $ over $ \Z_{q} $ having a monic generator polynomial $ g_{\alpha}(x) $ such that $g_{\alpha}(x) \mid x^{\alpha}-1$, and let $ C_{\beta} $ be principally generated free skew $ (\lambda_{0}+u\lambda_{1})$-constacyclic code of length $ \beta $ over $R$ with respect to the automorphism $\theta $ having a monic generator polynomial $ g_{\beta}(x)$ such that $ g_{\beta}(x) \mid x^{\beta}-(\lambda_{0}+u\lambda_{1})$. Then the code $ C $ generated by $ g(x)=(g_{\alpha}(x), g_{\beta}(x)) $ is a skew constacyclic code. Moreover, $\{g(x), xg(x), \ldots , x^{l-1}g(x) \}$ is a spanning set of $ C $, where $ l=deg (h(x)) $, $\eta(h(x))=h_{\alpha}(x)$, and $h_{\beta}(x)\mid h(x).$
\end{thm}

\pf
Let $ x^{\alpha}-1=h_{\alpha}(x)\ast g_{\alpha}(x) $ and $  x^{\beta}-(\lambda_{0}+u \lambda_{1})=h_{\beta}(x) \ast g_{\beta}(x)$,
for some monic polynomials $ h_{\alpha} \in \Z_{q}[x]$ and $ h_{\beta}(x) \in  R[x; \theta]$. Let $\rho(h(x))=h_{\alpha}(x)$ and $h_{\beta}(x)\mid h(x)$.\\
 Then
\[
 h(x) g(x)=h(x) (g_{\alpha}(x), g_{\beta}(x))=0
\]
as
\[
\eta(h(x))g_{\alpha}(x)= h_{\alpha}(x) \ast g_{\alpha}(x)=0,
\]
and
\[
 h(x)\ast g_{\beta}(x)=h'(x) \ast\ h_{\beta}(x) \ast g_{\beta}(x)=0.
\]

\noindent Now let $ v(x)\in C $, then $v(x)=r(x) g(x)$ for some $r(x)\in  R[x; \theta] $. We have
\[
r(x)=q(x) h(x)+t(x),
\]
\noindent where $t(x)=0  $ or $\deg(t(x)) < \deg(h(x))$. \\
Then

\[
 v(x)= r(x) g(x)=t(x) g(x).
\]
The result follows from the fact that  $t(x)=0  $ or $ \deg(t(x)) < \deg(h(x))$.  \qed

\subsection{The Gray Images of  Skew Constacyclic Codes over  $\mathbb{Z}_{q}R$}

We define a Gray map $\Phi$ from $\Z_{q}^{\alpha}R^{\beta}$ to $\Z^{\alpha+2\beta}_{q} $ by
\[
\begin{array}{c}
\Phi :\Z_{q}^{\alpha}R^{\beta} \rightarrow \Z^{\alpha+2\beta}_{q}\\
(e_{0}, \ldots , e_{\alpha-1}, r_{0}, \ldots , r_{\beta-1}) \mapsto (e_{0}, e_{1}, \ldots, e_{\alpha-1}, b_{0}, b_{1}, \ldots , b_{\beta-1}, a_{0}+b_{0}, a_{1}+b_{1}, \ldots , a_{\beta-1}+b_{\beta-1})
\end{array}
\]
where $r_{i}=a_{i}+ub_{i}$ for $i=0, 1, \ldots , \beta-1$.\\


\noindent {Let $ \lambda_{1}, \lambda_{2}, \ldots , \lambda_{l} $ be non-zero constants  in $\Z_{q}$. A linear code of length $ n = Nl  $ is  called a generalized QT  code, which is in turn a special case multi-twisted (MT) codes \cite{halil}, if for any codeword
\[
(c_{1}, c_{2}, \ldots , c_{(N-1)l}, c_{(N-1)l+1}, \ldots , c_{Nl}) \in C
\]
we have that
\[
(\lambda_{1}c_{Nl}, \lambda_{2}c_{Nl-1}, \ldots , \lambda_{l}c_{(N-1)l+1}, c_{1}, \ldots, c_{(N-1)l})\in C.
\]
\begin{thm}
Let $ C $ be a skew constacyclic code in $ \Z_{q}[x]/ \langle x^{\alpha}-1 \rangle  \times R[x; \theta]/ \langle x^{\beta}-(\lambda_{0}+u\lambda_{1})\rangle$, then the following hold.

(i) If $ \alpha=\beta $, then $ \Phi(C) $ is a $ (\lambda_{0}+u\lambda_{1})$-QT code with index $ 2 $ and length $3\alpha.$

(ii) If  $ \alpha \neq \beta $, then $ \Phi(C) $ is a generalized $(\lambda_{0}+u\lambda_{1})$-QT code with index $ 2$ and length $(\alpha, 2\beta) $.
\end{thm}

\pf
Let $C$ be a skew constacyclic code in $ \Z_{q}[x]/ \langle x^{\alpha}-1 \rangle  \times R[x; \theta]/ \langle x^{\beta}-(\lambda_{0}+u\lambda_{1})\rangle$ and let
\[
(e_{0}, e_{1}, \ldots, e_{\alpha-1}, b_{0}, b_{1}, \ldots , b_{\beta-1}, a_{0}+b_{0}, a_{1}+b_{1}, \ldots , a_{\beta-1}+b_{\beta-1})\in \Phi(C).
\]
Then there is a codeword $ (e_{0}, e_{1}, \ldots, e_{\alpha-1}, a_{0}+ub_{0}, a_{1}+ub_{1}, \ldots , a_{\beta-1}+ub_{\beta-1})\in C $ such that
\[
\Phi(e_{0}, e_{1}, \ldots , e_{\alpha-1}, r_{0}, r_{1}, \ldots , r_{\beta-1})= (e_{0}, e_{1}, \ldots, e_{\alpha-1}, b_{0}, b_{1}, \ldots , b_{\beta-1}, a_{0}+b_{0}, a_{1}+b_{1}, \ldots , a_{\beta-1}+b_{\beta-1}),
\]
where $ r_{i}=a_{i}+ub_{i} \textit{ for } i=0, 1, \ldots ,\beta-1.$\\
Since $ C$ is skew constacyclic, we have
\[
(e_{\alpha-1}, e_{0}, \ldots , e_{\alpha-2},  (\lambda_{0}+u\lambda_{1})\theta(r_{\beta-1}), \theta(r_{0}), \ldots , \theta(r_{\beta-2}))\in C
\]
and
\[
\begin{array}{ccl}
& &\Psi(e_{\alpha-1}, e_{0}, \ldots , e_{\alpha-2}, (\lambda_{0}+u\lambda_{1})\theta(a_{\beta-1}+ub_{\beta-1}), \theta(a_{0}+ub_{0}), \ldots , \theta(a_{\beta-2}+ub_{\beta-2}))\\
&=& \Psi(e_{\alpha-1}, e_{0}, \ldots , e_{\alpha-2},(\lambda_{0}+u\lambda_{1})(a_{\beta-1}+kb_{\beta-1}+udb_{\beta-1}), a_{0}+kb_{0}+udb_{0}, \ldots , a_{\beta-2}+kb_{\beta-2}+udb_{\beta-2})\\
&=&\Psi(e_{\alpha-1}, e_{0}, \ldots , e_{\alpha-2}, \lambda_{0}(a_{\beta-1}+kb_{\beta-1})+u(\lambda_{1}a_{\beta-1}+(\lambda_{0}d+\lambda_{1}k)b_{\beta-1}), a_{0}+kb_{0}+udb_{0}, \ldots ,\\
&& a_{\beta-2}+kb_{\beta-2}+udb_{\beta-2})\\
 &=&(e_{\alpha-1}, e_{0}, \ldots , e_{\alpha-2}, \lambda_{1}a_{\beta-1}+(\lambda_{0}d+\lambda_{1}k)b_{\beta-1},db_{0} \ldots , db_{\beta-2},
(\lambda_{0}+\lambda_{1})a_{\beta-1}+((k+d)\lambda_{0}+k\lambda_{1}) b_{\beta-1},\\
&& a_{0}+(k+d)b_{0},\ldots , a_{\beta-2}+(k+d)b_{\beta-2}).
\end{array}
\]
If $ \alpha=\beta$, then $ n=3\alpha $, thus $\Phi(C) $ is $ (\lambda_{0}+u\lambda_{1}) $-QT with index 2. Otherwise $\Phi(C)$ is a generalized $ (\lambda_{0}+u\lambda_{1})$-QT code with index 2 and length $(\alpha, 2\beta)$.
\qed
\subsection{Double Skew Constacyclic Codes over $\mathbb{Z}_{q}R$ }
In this subsection, we study double skew constacyclic codes over $\mathbb{Z}_{q}R$.
Let $n_{1}=\alpha+2\beta$ and $ n_{2}=\alpha'+2\beta'$ be integers such that $n = n_{1}+n_{2}$.
We consider a partition of the set of the $ n$ coordinates into two subsets of $n_{1}$ and $ n_{2} $ coordinates, respectively, so that $ C $ is a subset of $\Z_{q}^{\alpha}R^{\beta} \times \Z_{q}^{\alpha'}R^{\beta'}$.
\begin{defi}
 A linear code $C$ of length $ n $ over $ \Z_{q}R $ is called a double skew constacyclic code if $C$ satisfies the following conditions.

(i) $ C$ is a linear code.

(ii) If
\[
( e_{0}, e_{1}, \ldots , e_{\alpha-1}, r_{0}, r_{1}, \ldots , r_{\beta-1}, e'_{0}, e'_{1}, \ldots , e'_{\alpha'-1}, r'_{0}, r'_{1}, \ldots , r'_{\beta'-1}) \in C
\]
then
\[
\begin{array}{cccc}
 (  e_{\alpha-1}
, \ldots , e_{\alpha-2}, \lambda\theta( r_{\beta-1})
, \ldots ,\theta ( r_{\beta-2}),
 e'_{\alpha'-1}
, \ldots , e'_{\alpha'-2}, \lambda\theta( r'_{\beta'-1})
, \ldots ,\theta ( r'_{\beta'-2}) )  \in C,
\end{array}
\]
where $ \lambda= \lambda_{0}+u\lambda_{1} $.
\end{defi}
 Denote by $ \mathfrak{R}_{\alpha, \beta, \alpha', \beta'} $ the ring $\Z_{q}[x ]/ \langle x^{\alpha}-1 \rangle  \times R[x; \theta]/ \langle x^{\beta}-(\lambda_{0}+u\lambda_{1})\rangle \times  \Z_{q}[x]/ \langle x^{\alpha'}-1 \rangle  \times R[x; \theta]/ \langle x^{\beta'}-(\lambda_{0}+u\lambda_{1})\rangle.$\\
In polynomial representation, each codeword
\[
c= (e_{0}, e_{1}, \ldots, e_{\alpha-1}, r_{0}, r_{1}, \ldots, r_{\beta-1}, e'_{0}, e_{1}, \ldots , e'_{\alpha'-1}, r'_{0}, r'_{1}, \ldots, r'_{\beta'-1} )
\]
  of a skew constacyclic code can be represented by four polynomials

 \begin{equation*}
c(x)=\left(
\begin{array}{c}
e_{0}+ e_{1}x+ \ldots + e_{\alpha-1}x^{\alpha-1},\\
r_{0}+ r_{1}x+ \ldots +r_{\beta-1}x^{\beta-1},\\
e'_{0}+ e'_{1}x+ \ldots + e'_{\alpha'-1}x^{\alpha'-1},\\
r'_{0}+ r'_{1}x+ \ldots +r'_{\beta'-1}x^{\beta'-1}
\end{array}
\right)
=(e(x), r(x), e'(x), r'(x))\in   \mathfrak{R}_{\alpha, \beta, \alpha', \beta'} .
\end{equation*}
Let
\[
h(x)=h_{0}+ h_{1}x+ \ldots + h_{t}x^{t} \in R \ [ x; \theta]
\]
and let
 \[
   (f(x), g(x), f'(x), g'(x)  )\in  \mathfrak{R}_{\alpha, \beta, \alpha', \beta'}.
\]
 We define a multiplication by
\[
h(x) (f(x), g(x) )=(\eta(h(x))f(x), h(x) \ast g(x), \eta(h(x))f'(x), h(x) \ast g'(x) ),
\]
where $\eta(h(x))=\eta( h_{0})+\eta( h_{1})x+ \ldots +\eta( h_{t})x^{t}.$ This gives us the following Theorem.\\
\begin{thm}
A linear code $ C $ is a double skew constacyclic code if and only if it is a  left $ R[x; \theta]$-submodule of $ R_{\alpha, \beta, \alpha', \beta'}$.
\end{thm}
\pf
Let $ c(x) =c_{0}+c_{1}x+ \ldots + c_{n-1}x^{n-1}$ be a codeword of $ C$. Since $C$ is a double skew $( \lambda_{0}+u\lambda_{1})$-constacyclic code, it follows that $ x c(x) \in  C$. By linearity of $  C$, $ r(x) c(x) \in  C $ for any $ r(x) \in R[x; \theta]$. Therefore $  C $ is a left $ R[x; \theta]$-submodule of $ \mathfrak{R}_{\alpha, \beta, \alpha', \beta'}$.
Conversely, suppose that $ C$ is a left $R[x; \theta]$-submodule of $  \mathfrak{R}_{\alpha, \beta, \alpha', \beta'}$, then we have that $x  c(x) \in C$. Thus, $C$ is a double skew $(\lambda_{0}+u\lambda_{1})$-constacyclic code.
\qed

\section{New Linear Codes over $\mathbb{Z}_4$}
Codes over $\mathbb{Z}_4$, sometimes called quaternary codes as well, have a special place in coding theory. Due to their importance, a database of quaternary codes was introduced in \cite{Z4DBpaper} and it is available online  \cite{Z4db}. Hence we consider the case $q=4$ to possibly obtain quaternary codes with good parameters. We conducted a computer search using Magma software \cite{magma} to find skew cyclic codes over $\mathbb{Z}_{4}(\mathbb{Z}_{4}+u\mathbb{Z}_{4})$ whose Gray images are quaternary linear codes with better parameters than the currently best known codes. We have found ten such codes which are listed in the table below.

The automorphism of $R=\mathbb{Z}_{4}+u\mathbb{Z}_{4}$ that we used is $\theta(a+bu)=a+3bu=a-bu$. In addition to the Gray map given in Section 4.1, there are many other possible linear maps from $\mathbb{Z}_{4}+u\mathbb{Z}_{4}$  to $\mathbb{Z}_{4}^{\ell}$ for various values of $\ell$. For example, the following map was used in \cite{constaZ4uZ4} $a+bu\to (b,2a+3b,a+3b)$ which triples the length of the code. We used both of these Gray maps in our computations, and obtained new codes from each map.

We first chose a cyclic code $C_{\alpha}$ over $\mathbb{Z}_4$ generated by $g_{\alpha}(x)$. The coefficients of this polynomial is given in ascending order of the terms in the table. Therefore,  the entry 31212201, for example, represents the polynomial $ 3+x+2x^2+x^3+2x^4+2x^5+x^7$. Then we searched for divisors of $x^{\beta}-1$ in the skew polynomial ring $R[x;\theta]$ where $R=\mathbb{Z}_{4}+u\mathbb{Z}_{4}$ and $\theta(a+bu)=a-bu$. For each such divisor $g_{\beta}(x)$ we constructed the skew cyclic code over  $\mathbb{Z}_{4}R$ generated by $(g_{\alpha}(x),g_{\beta}(x))$ and its $\mathbb{Z}_4$-images under each Gray map described above. As a result of the search, we obtained ten new linear codes over $\mathbb{Z}_4$. They are now added to the database (\cite{Z4db}) of quaternary codes. In Table 1 below, which Gray map is used to obtain each new code is not explicitly stated, but it can be inferred from the values of $\alpha, \beta$ and $n$, the length of the  $\mathbb{Z}_4$ image. If $n=\alpha+2\beta$, then it is the map given in section 4.1 and if $n=\alpha+3\beta$ it is the map described in this section. For example, the second code in the table has length $57=15+3\cdot 14$. This means that the Gray map that triples the length of a code over $R$ is used to obtain this code. 

When $x^{\beta}-1=g(x)\ast h(x)$ we can use either the generator polynomial $g(x)$ or the parity check polynomial $h(x)$ to define the skew cyclic code over $R$. For the codes given in the table below we used the parity check polynomial because it has smaller degree. In general a linear code $C$ over $\mathbb{Z}_{4}$ has parameters
$[n,4^{k_1}2^{k_2}]$, and when $k_2=0$, $C$ is a free code. In this case $C$ has a basis with $k$ vectors just like a linear code over a field. All of the codes in the table below are free codes, hence we will simply denote their parameters by $[n,k,d]$ where $d$ is the Lee weight over $\mathbb{Z}_4$. 

Our computational results suggest that considering skew cyclic and skew constacyclic codes over  $\mathbb{Z}_{q}(\mathbb{Z}_{q}+u\mathbb{Z}_{q})$ is promising to obtain codes with good parameters over $\mathbb{Z}_{q}$.

\begin{table}[!htbp]
 \noindent\caption{New  quaternary  codes }
\begin{center}
\begin{tabular}
[c]{|c|c|c|c|c|} \hline \textbf{$\mathbf{\alpha}$} & \ \ \  \textbf{$\mathbf{\beta}$} & \ \ \  $\mathbf{g_{\alpha}}$  & \ \ \   $\mathbf{h_{\beta}}$\ \ \ &  \textbf{ {$\mathbb{Z}_4$}  Parameters}
\\\hline
$15$ & $14$ & $ 31212201   $ &  $ x^4 + (u + 1)x^3 + x^2 + (3u + 2)x + 3u + 3$ & $[43,8,26]$ \\\hline
$15$ & $14$ & $ 31212201   $ &  $ x^4 + (u + 1)x^3 + x^2 + (3u + 2)x + 3u + 3$ & $[57,8,38]$ \\\hline
$15$ & $14$ & $ 3021310231   $ &  $ x^3+2ux^2 + (3u + 3)x + 2u+ 3$ & $[43,6,30]$ \\\hline
$15$ & $14$ & $ 3021310231   $ &  $ x^3+ 3x^2+(3u+2)x+1$ & $[57,6,42]$ \\\hline
$7$ & $14$ & $ 3121 $ &  $ x^4+(3u+3)x^3+3x^2+(u+2)x+3u+3$ & $[35,8,20]$ \\\hline
$7$ & $14$ & $ 3121 $ &  $ x^4+(u+3)x^3+(u+1)x^2+(u+2)x+3u+3$ & $[49,8,32]$ \\\hline
$7$ & $14$ & $ 12311 $ &  $ x^3+(2u+1)x^2 + 3ux + 3u + 3$ & $[35,6,22]$ \\\hline
$7$ & $14$ & $12311 $ &  $ x^3+(2u+1)x^2+ux+u+1$ & $[35,6,24]$ \\\hline
$7$ & $14$ & $ 12311 $ &  $ x^3+ux^2+(3u+3)x+1$ & $[49,6,35]$ \\\hline
$7$ & $14$ & $ 12311 $ &  $ x^3+(u+2)x^2+x+1$ & $[49,6,36]$ \\\hline

\end{tabular}
\end{center}
\end{table}

\section{Conclusion}
In this paper skew constacyclic codes are considered over the ring $\Z_{q}R$, where $ R=\mathbb{Z}_{q}+u\mathbb{Z}_{q}$, $q$ is a prime power and $u^{2}=0$ and their algebraic and structural properties are studied. Considering
their Gray images, we obtained some new linear codes over $\mathbb{Z}_4$ from skew cyclic codes over $\Z_{q}R$. Moreover, these codes are then generalized to double skew constacyclic codes.


\begin{thebibliography}{99}


\bibitem{abua} T. Abualrub, I. Siap, \textit{Cyclic codes over the rings  $  \mbox{\msbm Z}_{2}+u \mbox{\msbm Z}_{2} $ and $  \mbox{\msbm Z}_{2}+u \mbox{\msbm Z}_{2} +u^{2} \mbox{\msbm Z}_{2}$}, Designs, Codes and Cryptography, 42 (3) (2007), pp. 273--287.

\bibitem{taher} T. Abualrub, I. Siap and I. Aydogdu, \textit{$ \mbox{\msbm Z}_{2} (\mbox{\msbm Z}_{2}+u\mbox{\msbm Z}_{2})$-Linear cyclic codes}, Proceedings of the IMECS 2014, 2 (2014), Hong Kong.

\bibitem{qc4}  R. Ackerman, N. Aydin, New quinary linear codes from quasi-twisted codes and their duals, Appl. Math. Lett.,  24(4) (2011), pp. 512--515.

\bibitem{ayats} J. B. Ayats, C. F. C\'{o}rdoba and R. T. Valls, \textit{$ \mbox{\msbm Z}_{2} \mbox{\msbm Z}_{4}$-additive cyclic codes, generator polynomials and dual codes}, IEEE Transactions on Information Theory, 62 (2016), pp. 6348--6354.

\bibitem{ismail} I. Aydogdu, T. Abualrub and I. Siap, \textit{$ \mbox{\msbm Z}_{2}\mbox{\msbm Z}_{2}[u] -$cyclic and constacyclic codes}, IEEE Transactions on Information Theory, 63 (8) (2016), pp. 4883--4893.

\bibitem{Z4DBpaper} N. Aydin, T. Asamov, \textit{A Database of $\mathbb{Z}_4$ Codes},  Journal of Combinatorics, Information \& System Sciences, 34 (1-4) (2009), pp. 1--12.

\bibitem{qc6} N. Aydin, N. Connolly and M. Grassl, \textit{Some results on the structure of constacyclic codes and new linear codes over $GF(7)$ from quasi-twisted codes}, Adv.  Math. of Commun., 11 (1)(2017), pp. 245--258.

\bibitem{qc5} N. Aydin, N. Connolly and J. Murphree, \textit{New binary linear codes from QC codes and an augmentation algorithm}, Appl. Algebra Eng. Commun. Comput., 28( 4) (2017), pp. 339--350.


\bibitem{constaZ4uZ4} N. Aydin, Y. Cengellenmis and A. Dertli, \textit{On some constacyclic codes over $\mathbb{Z}_4[u]/ \langle u^2-1 \rangle$, their  $ \mathbb{Z}_4 $ images, and new codes}, Designs, Codes and Cryptography, 86 (6) (2018), pp. 1249--1255.

\bibitem{qtmain} N. Aydin, I. Siap and D. Ray-Chaudhuri, \textit{The structure of 1-generator quasi-twisted codes and new linear codes}, Designs, Codes and Cryptography, 24 (3) (2001), pp. 313--326.

\bibitem{qc1}  N. Aydin, I. Siap, \textit{New quasi-cyclic codes over $\mathbb{F}_5$}, Appl. Math. Lett., 15 (7) (2002), pp. 833--836.

\bibitem{halil}  N. Aydin, A. Halilovi\'{c}, \textit{A Generalization of Quasi-twisted Codes: Multi-twisted codes}, Finite Fields and Their Applications, 45 (2017), pp. 96--106.

\bibitem{rama} R. K. Bandi, M. Bhaintwal, \textit{A note on cyclic codes over $ \mbox{\msbm Z}_{4}+u\mbox{\msbm Z}_{4} $}, Discrete Mathematics, Algorithms and Applications, 8 (1) (2016), pp. 1--17.

\bibitem{nabil} N. Bennenni, K. Guenda and S. Mesnager, \textit{DNA cyclic codes over rings}, Adv. in Math. of Comm., 11 (1) (2017), pp. 83--98.

\bibitem{skew1} D. Boucher, W. Geiselmann and F. Ulmer, \textit{Skew-cyclic codes}, Appl. Algebra Engrg. Comm. Comput., 18(4)(2007), pp. 379--389.


\bibitem{qc2} R. Daskalov, P. Hristov, \textit{New binary one-generator quasi-cyclic  codes},  IEEE Trans.  Inf. Theory,
 49 (11) (2003), pp 3001--3005.

\bibitem{qc3} R. Daskalov, P. Hristov and E. Metodieva, \textit{New minimum distance bounds for linear codes over GF(5)}, Discrete Math., 275 (1--3) (2004), pp. 97--110.


\bibitem{Z4db}  Database of $\mathbb{Z}_4$ Codes. [online] Z4Codes. info (Accessed March, 2018).



\bibitem{abhay} H. Q. Dinh, A. K. Singh, S. Pattanayak and S. Sriboonchitta, \textit{ Cyclic DNA codes over the ring $\mathbb{F}_2+u \mathbb{F}_2+v \mathbb{F}_2+uv \mathbb{F}_2+v^2 \mathbb{F}_2+uv^2 \mathbb{F}_2$}, Designs, Codes and Cryptography, 86 (7) (2018), pp. 1451-1467.

\bibitem{ezerman}   M.F. Ezerman, S. Ling, P. Sol\'e and O. Yemen, \textit{From skew-cyclic codes to asymmetric quantum code}, Adv. in Math. of Comm., 5 (1) (2011), pp. 41--57.
\bibitem{gao} J. Gao. \textit{Skew cyclic codes over $\mathbb{F}_{p}+v\mathbb{F}_{p}$}. J. Appl. Math. Inform., 31 (3--4)(2013), pp. 337--342.

\bibitem{fu} J. Gao, F. W. Fu, L. Xiao and R. K. Bandi, \textit{Some results on cyclic codes over $ \mbox{\msbm Z}_{q} +u\mbox{\msbm Z}_{q}$}, Discrete Mathematics, Algorithms and Applications, 7 (4) (2015), pp. 1--9.

\bibitem{jian} J. Gao, F. Ma and F. Fu, \textit{Skew constacyclic codes over the ring $\mathbb{F}_{q}+v\mathbb{F}_{q},$} Appl.Comput. Math., 6 (3) (2017), pp. 286--295.

\bibitem{database} M. Grassl, \textit{Code Tables: Bounds on the parameters of codes}, online, http://www.codetables.de/


\bibitem{gursoy} F. Gursoy, I. Siap and B. Yildiz, \textit{Construction of skew cyclic codes over $ \mbox{\msbm F}_{q}+v\mbox{\msbm F}_{q} $}, Advances in Mathematics of Communications, 8 (3) (2014), pp. 313--322.

\bibitem{jit} S. Jitman, S. Ling and P. Udomkavanich, \textit{Skew constacyclic over finite chain rings}, Adv. Math.
Commun., 6 (1) (2012), pp. 39--63.

\bibitem{li} P. Li, W. Dai and X. Kai, \textit{On $ \mbox{\msbm Z}_{2}\mbox{\msbm Z}_{2}[u]-(1 + u)$-additive constacyclic}, arXiv:1611.03169v1  [cs.IT] 10 Nov 2016.

\bibitem{magma} Magma computer algebra system, online, http://magma.maths.usyd.edu.au/

\bibitem{qian} J. F. Qian, L. N. Zhang and S. X. Zhu, \textit{$ (1+u) $-Constacyclic and cyclic codes over $ \mbox{\msbm F}_{2}+u\mbox{\msbm F}_{2} $}, Applied Mathematics Letters, 19 (8) (2006), pp. 820--823.

\bibitem{amit} A. Sharma, M. Bhaintwal, \textit{A class of skew-constacyclic codes over $ \mbox{\msbm Z}_{4}+u\mbox{\msbm Z}_{4} $}, Int. J. Information and Coding Theory, 4 (4)(2017), pp. 289--303.



\bibitem{siap} I. Siap, T. Abualrub, N. Aydin and P. Seneviratne, \textit{Skew cyclic codes of arbitrary length},
Int. J. Information and Coding Theory, 2 (1) (2011), pp. 10--20.

\bibitem{yildiz} B. Yildiz, N. Aydin, \textit{Cyclic codes over $  \mbox{\msbm Z}_{4}+u \mbox{\msbm Z}_{4}$ and their $\mbox{\msbm Z}_{4}  $-images}, Int. J. Information and coding Theory, 2 (4) (2014), pp. 226--237.

\end{thebibliography}
\end{document}